\documentclass[cits]{PoS}
\usepackage{amssymb}

\title{Anisotropies in the diffuse gamma-ray background measured by the Fermi-LAT }

\ShortTitle{Anisotropies in the Fermi gamma-ray background}

\author{\speaker{Jennifer M.~SIEGAL-GASKINS}\\%
       CCAPP, The Ohio State University\\
       E-mail: \email{jsg@mps.ohio-state.edu}}

\author{on behalf of the \emph{Fermi}-LAT Collaboration and Eiichiro Komatsu}

\abstract{The contribution of unresolved sources to the diffuse gamma-ray background could produce anisotropies in this emission on small angular scales. Recent studies have considered the angular power spectrum and other anisotropy metrics as tools for identifying contributions to diffuse emission from unresolved source classes, such as extragalactic and Galactic dark matter as well as various astrophysical gamma-ray source populations. We present preliminary results of an anisotropy analysis of the diffuse emission measured by the \emph{Fermi-}LAT.}

\FullConference{Identification of Dark Matter 2010\\
		 July 26 - 30 2010\\
		 University of Montpellier 2, Montpellier, France}

\begin{document}

\section{Introduction}

The Large Area Telescope (LAT) on board the Fermi Gamma-ray Space Telescope (\emph{Fermi})~\cite{Atwood:2009ez} is observing the high-energy sky with unprecedented precision and sensitivity.  In addition to numerous individual sources, the LAT observes a substantial level of diffuse gamma-ray emission.  This emission includes a Galactic component resulting from the interactions of cosmic rays with the interstellar gas and radiation fields, as well as a component that appears isotropic on large angular scales, which is often assumed to originate from unresolved members of cosmological source populations.  

Many astrophysical sources are guaranteed to contribute to the large-scale isotropic gamma-ray background (IGRB), including cosmological populations such as blazars~\cite{Stecker:1996ma, Narumoto:2006qg} and star-forming galaxies~\cite{Thompson:2006np, Fields:2010bw}, as well as Galactic source classes whose sky distributions extend to high Galactic latitudes, e.g., millisecond pulsars~\cite{FaucherGiguere:2009df}.  Proposed but unconfirmed sources of gamma-ray emission, such as the annihilation or decay of dark matter particles in Galactic or extragalactic structures, may also contribute to the IGRB~\cite{Ullio:2002pj, Zavala:2009zr, Abdo:2010dk}.   
Interestingly, the \emph{Fermi}-measured IGRB energy spectrum~\cite{Abdo:2010nz} is consistent with a single power law over a large range of energies, and hence lacks any spectral features which could aid in identifying contributions from individual source classes.
Moreover, recent \emph{Fermi} results~\cite{Collaboration:2010gqa} indicate that the majority of the IGRB does not originate from members of the source classes already detected by \emph{Fermi}, leaving the origin of this emission a mystery.  

To complement searches based on spectral features, recent work has considered the possibility of using angular anisotropy information in the IGRB to help to reveal its contributors~\cite{Zavala:2009zr, Ando:2005xg, Ando:2006cr, SiegalGaskins:2008ge, Fornasa:2009qh, Ando:2009fp, Ando:2009nk}.
If the IGRB is composed of emission from unresolved members of gamma-ray source populations, characteristic small-scale fluctuations are expected to be present due to the variation in the number density of sources along the line-of-sight in different sky directions.  

In this study we searched for angular anisotropies in the IGRB measured by the \emph{Fermi}-LAT\@.  The angular power spectrum of the emission, after masking low Galactic latitudes and known sources, was calculated in several energy bins.  The results from the data were compared with those from a simulated model of the gamma-ray sky in order to identify any significant differences in anisotropy properties.  Preliminary results from this angular power spectrum analysis of the IGRB are presented.

\section{Data selections and processing}

The \emph{Fermi}-LAT is a pair-conversion telescope that uses a tracker and calorimeter to reconstruct the direction and energy of individual events.   An anti-coincidence detector and a tailored event classification scheme provide the LAT with excellent charged particle background rejection capabilities.  The LAT detects photons with energies from 20 MeV to more than 300 GeV, and achieves an angular resolution of approximately 0.1 deg for photons above 10 GeV\@.  The LAT has a large field of view $\Omega_{\rm fov} \sim 2.4$ sr, and operates primarily in sky-scanning mode, resulting in fairly uniform sky exposure of roughly 30 minutes every 3 hours.  Further details about the instrument can be found in Ref.~\cite{Atwood:2009ez}.

The anisotropy analysis was performed on $\sim$ 22 months of data, using ``diffuse class'' events in the energy range of 1 GeV to 50 GeV.  The events were binned into 5 energy ranges for the angular power spectrum calculation.   Using several energy bins increases the sensitivity of the analysis to source populations that only contribute significantly to the anisotropy in a limited energy range.  In addition, an observation of energy-dependent anisotropy may aid in interpretation of a measurement in terms of a detection of or constraints on specific source populations~\cite{SiegalGaskins:2009ux, Hensley:2009gh, Cuoco:2010jb}.  

To serve as a comparison, a simulated model of the gamma-ray sky was constructed using the {\tt gtobssim} module of the \emph{Fermi} Science Tools\footnote{\tt http://fermi.gsfc.nasa.gov/ssc/data/analysis/}.  The model was composed of a representation of the Galactic diffuse emission ({\tt gll\_iem\_v02.fit}, the publicly-available diffuse model recommended for LAT data analysis\footnote{{\tt http://fermi.gsfc.nasa.gov/ssc/data/access/lat/BackgroundModels.html}}), an isotropic component, and sources in the 11-month \emph{Fermi}-LAT source catalog~\cite{Collaboration:2010ru}.

The data and simulations were processed with the \emph{Fermi} Science Tools using the P6\_V3 instrument response functions.  The events were binned into order 9 HEALPix~\cite{Gorski:2004by} maps, corresponding to pixels of $\sim$ 0.1 deg/side.  At this map resolution the suppression of angular power due to the pixelation of the map (the pixel window function $W_{\rm pix}$) is subdominant compared to the suppression of angular power due to the point spread function (PSF) of the instrument (the beam window function $W_{\rm beam}$).  \emph{Fermi's} PSF for back-converting events is significantly poorer than for front-converting events, so the front- and back-converting events were analyzed separately through the angular power spectrum calculation.  This choice allowed for a more accurate estimation of the measurement uncertainties, which depend strongly on the PSF\@.

The goal of the analysis was to measure the angular power spectrum of the IGRB, so regions of the sky heavily contaminated by Galactic diffuse emission were excluded by masking Galactic latitudes $|b| < 30^{\circ}$, and masking sources in the \emph{Fermi} 11-month catalog~\cite{Collaboration:2010ru} within a $2^{\circ}$ angular radius.  In this study we focused on multipoles $\ell \gtrsim 100$ (corresponding to angular scales $\lesssim 2^{\circ}$), since lower multipoles (corresponding to correlations over larger angular scales) are likely more contaminated by Galactic diffuse emission.

We considered the angular power spectrum $C_{\ell}$ of a map of intensity fluctuations $\delta I$, with $\delta I (\psi) = (I(\psi) - \langle I \rangle)/\langle I \rangle$, where $I(\psi)$ is the intensity in the sky direction $\psi$ and $\langle I \rangle$ is the mean intensity of the unmasked region of the map.
The angular power spectrum is given by the coefficients $C_{\ell} = \langle |a_{\ell m}|^{2} \rangle$ with the $a_{\ell m}$ determined by expanding the map in spherical harmonics.  The angular power spectrum of intensity fluctuations is dimensionless and hence characterizes the angular distribution of the emission independent of the intensity normalization.  Using this convention, the amplitude of the angular power spectrum for a single source class is the same in all energy bins (if the source distribution is independent of energy).  However, we emphasize that if the IGRB is composed of emission from multiple source classes, the amplitude of the fluctuation angular power spectrum does not indicate the relative contribution of a single source class to the anisotropy of the total emission, i.e., the fluctuation angular power spectra from multiple source classes are not linearly additive.

The angular power spectra of the maps were calculated using the HEALPix package \cite{Gorski:2004by}.  The measured angular power spectra were corrected for the power suppression due to the beam and pixel window functions, and an approximate correction, valid at multipoles $\ell \gtrsim 100$, was applied to account for the reduction in angular power due to masking.  For each energy bin, the angular power spectra of the maps of front- and back-converting events were calculated separately and then combined by weighted average.

\section{Results}
\label{sec:results}

\begin{figure*}
\centering
\includegraphics[width=0.45\textwidth]{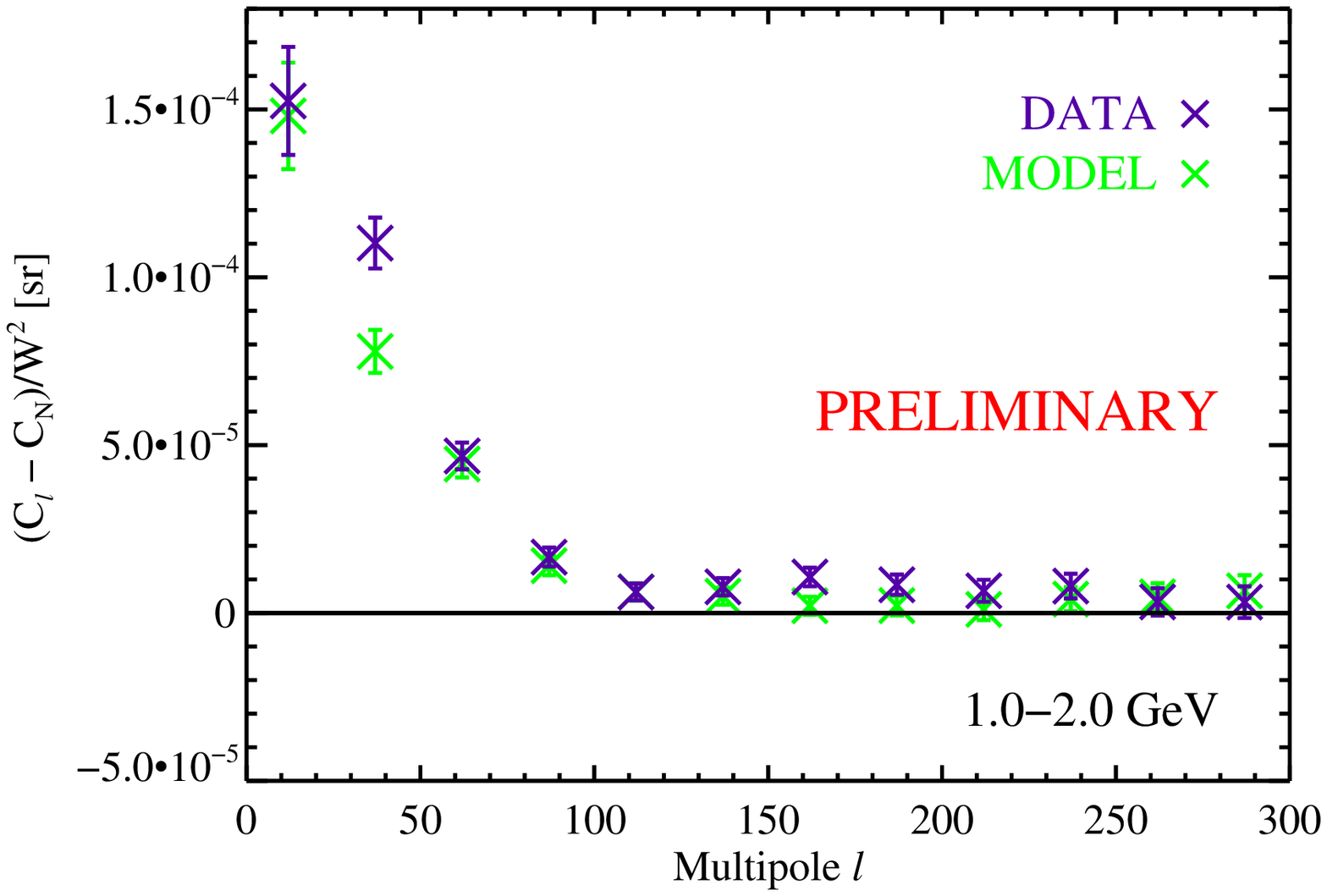}
\includegraphics[width=0.45\textwidth]{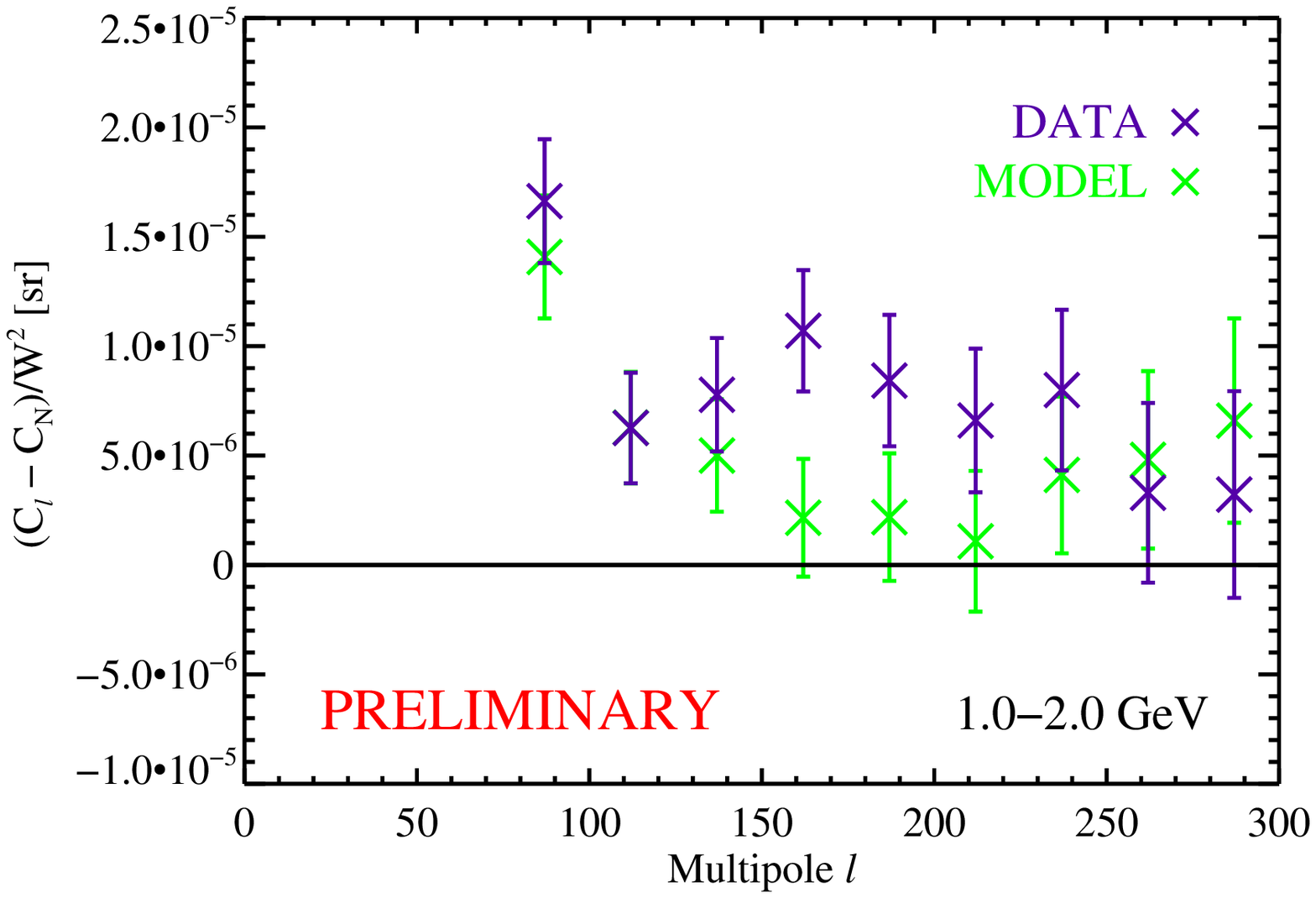}
\includegraphics[width=0.45\textwidth]{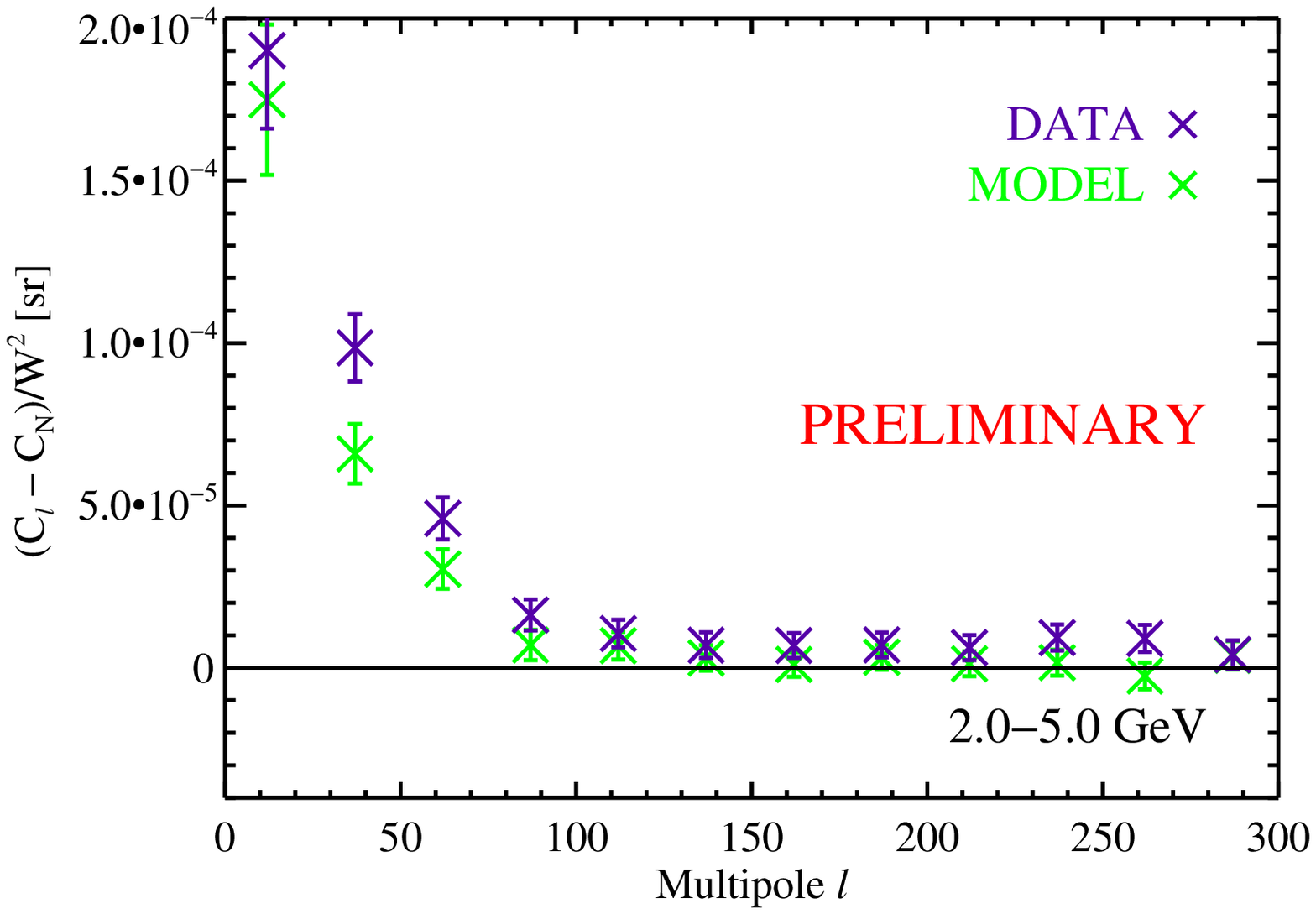}
\includegraphics[width=0.45\textwidth]{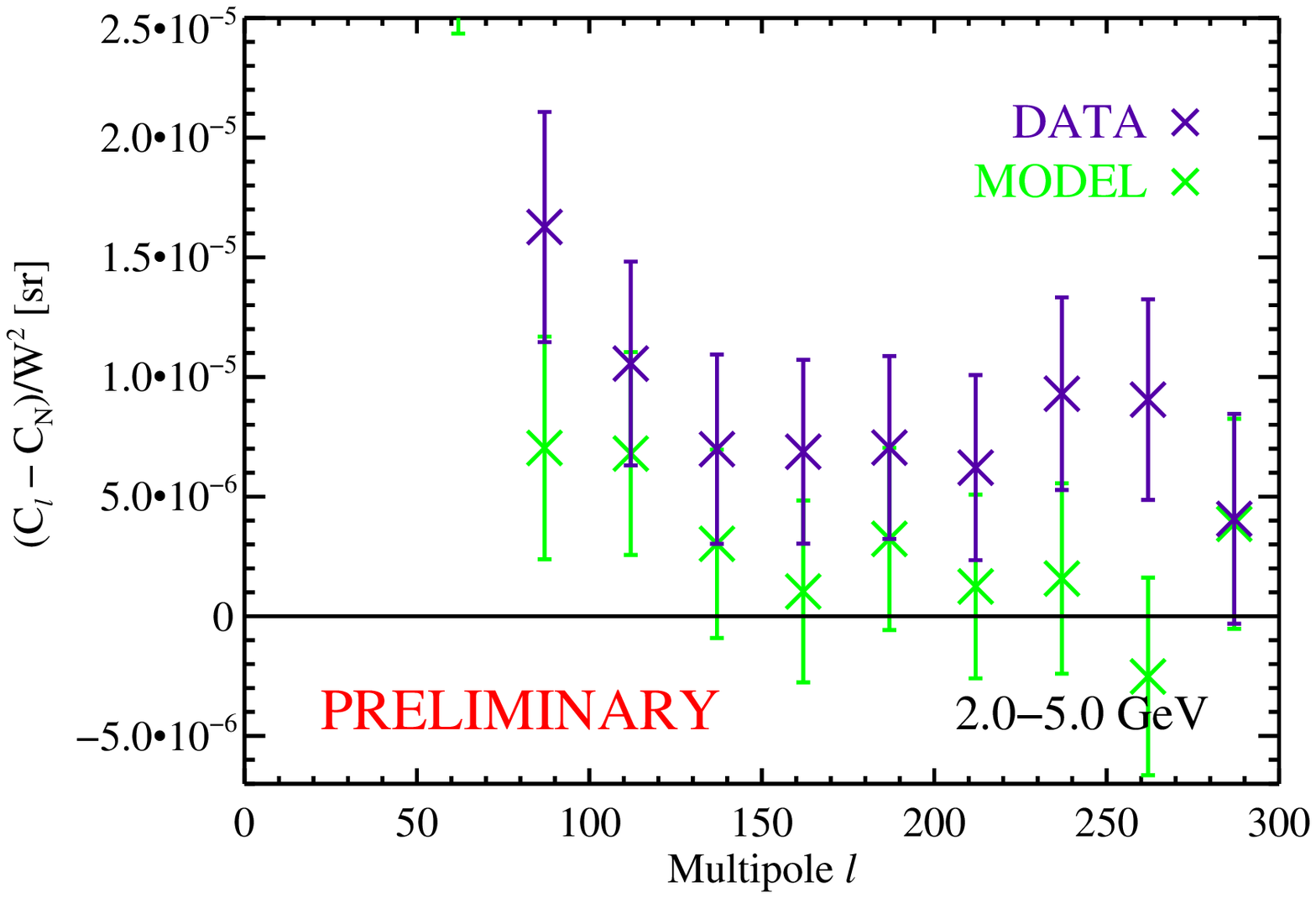}
\includegraphics[width=0.45\textwidth]{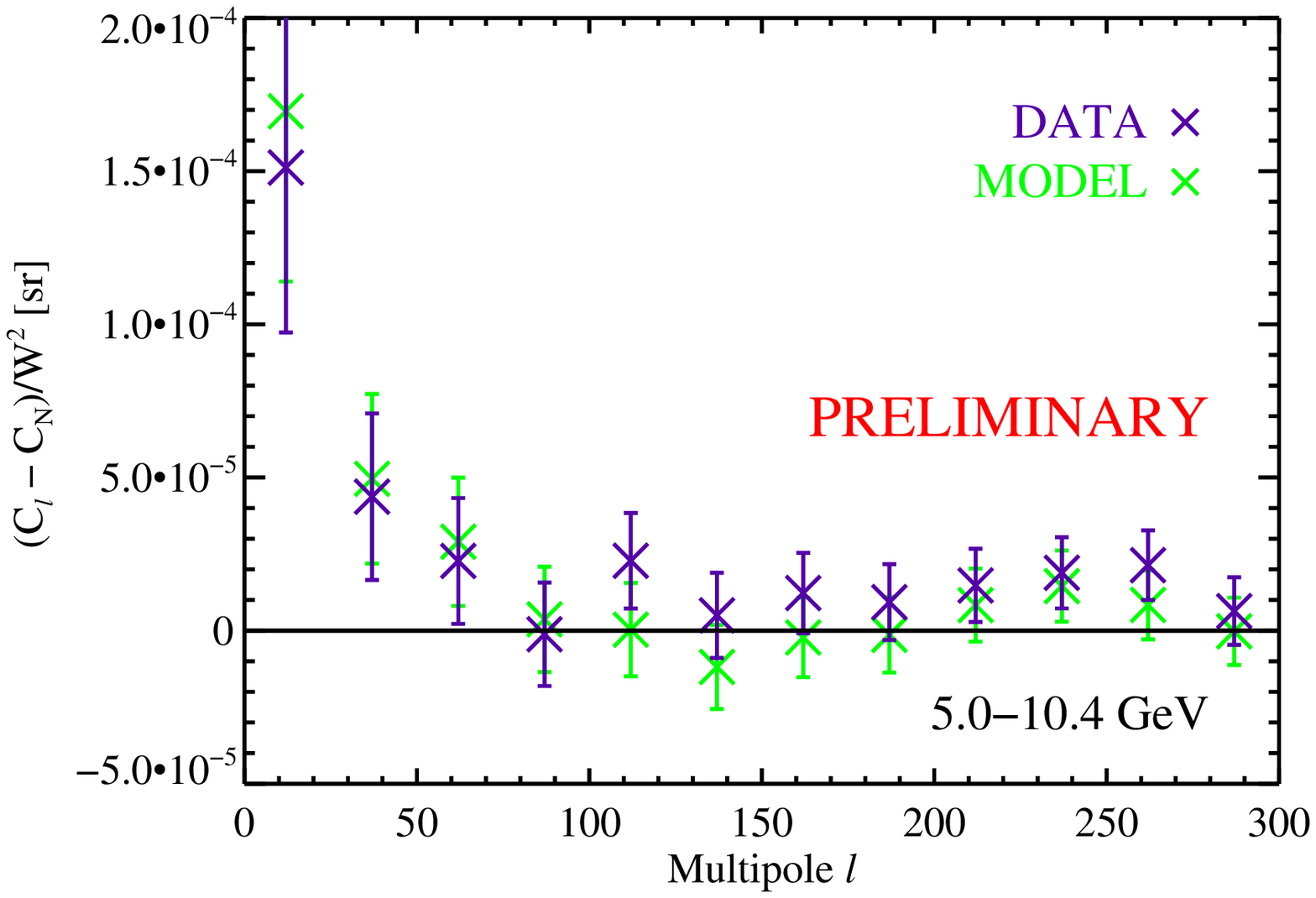}
\includegraphics[width=0.45\textwidth]{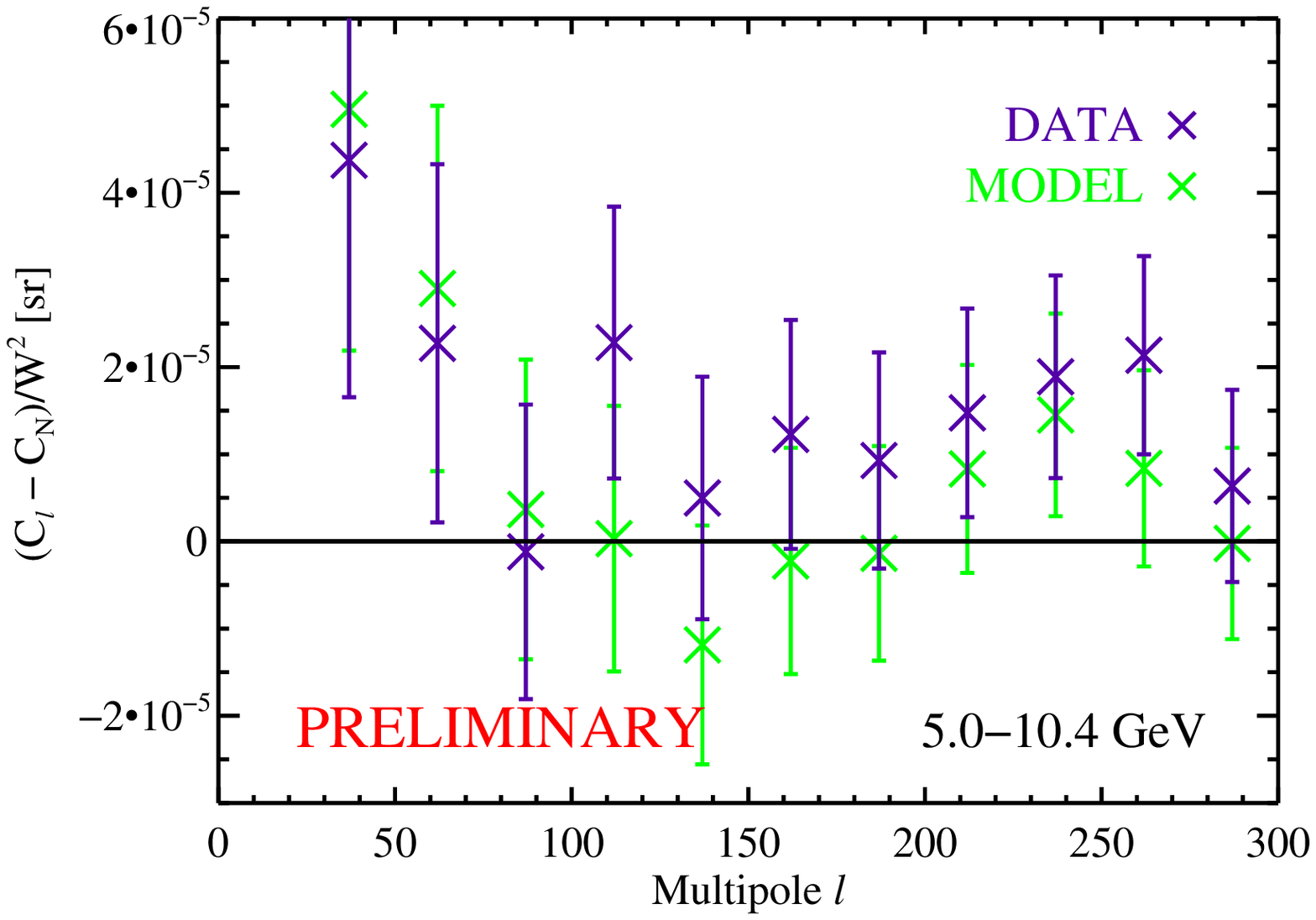}
\includegraphics[width=0.45\textwidth]{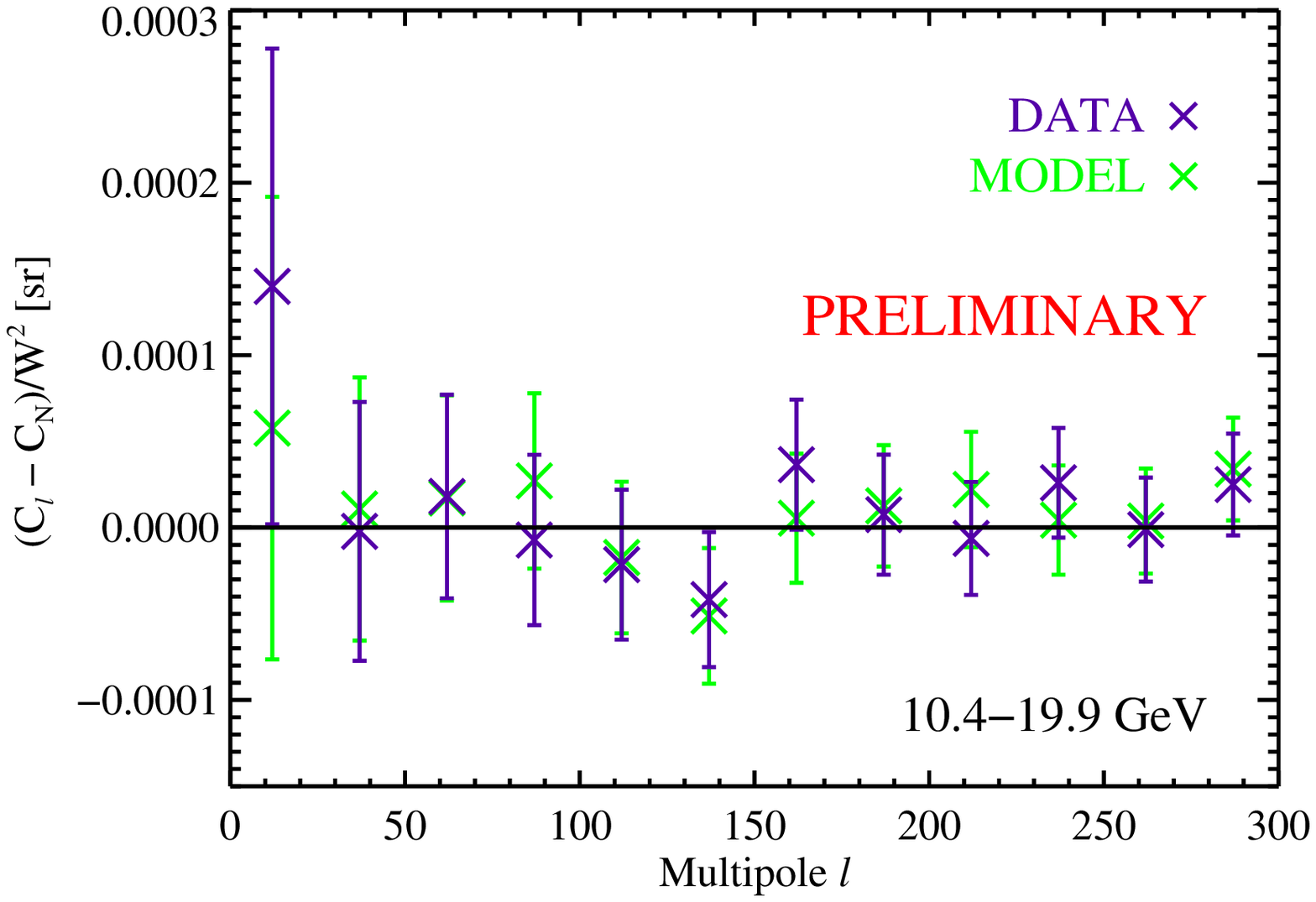}
\includegraphics[width=0.45\textwidth]{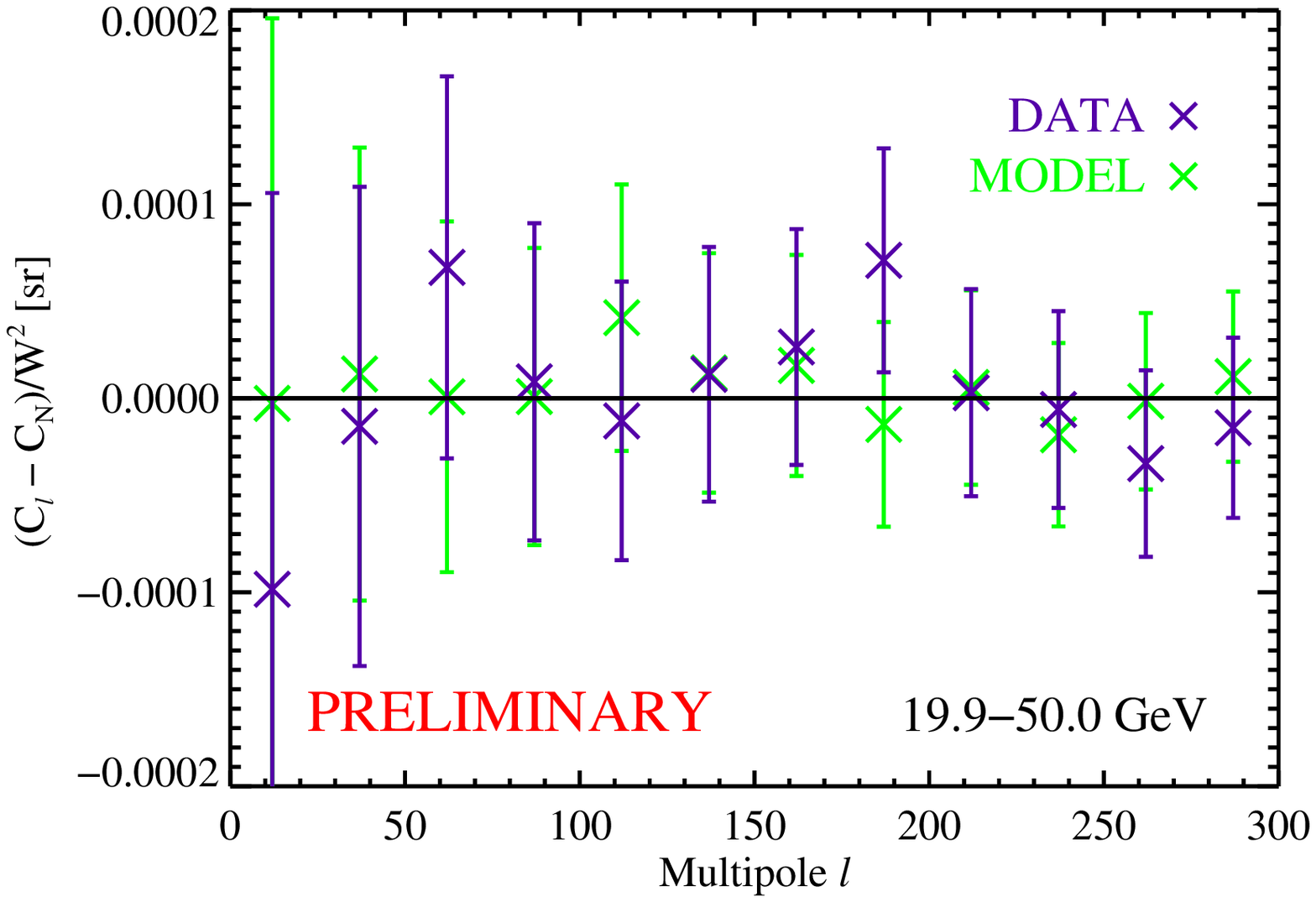}
\caption{Angular power spectra of \emph{Fermi}-LAT data and a simulated model in several energy bins.  Angular power spectra are shown with the photon noise level $C_{\rm N}$ subtracted, and have been corrected for the power suppression due to the beam and pixel window functions $W_{\ell}$. In the top three rows, the right panel shows the same result as the left panel, i.e., the angular power spectra for the same energy bin, but over a smaller range of $C_{\ell}$ to more clearly illustrate the result at multipoles $\ell \gtrsim 100$.\label{fig:aps}}
\end{figure*}

The angular power spectra of the data and of the model in several energy bins are shown in Fig.~\ref{fig:aps}.  The $C_{\ell}$ were averaged in multipole bins of width $\Delta \ell = 25$, and the angular power spectra are shown with the photon noise level $C_{\rm N}$ subtracted, so a measurement above zero indicates the presence of angular power above the noise level.  The error bars indicate the 1-$\sigma$ statistical uncertainty in the measurement; systematic uncertainties are not included.
We emphasize that uncertainties in the determination of the PSF can affect the calculation of the measurement uncertainties and the correction for the beam window function.

At multipoles $\ell \gtrsim 100$, angular power above the photon noise level is measured in the data at energies from 1 to 5 GeV; excess power in the data is also found at low significance in the 5-10 GeV energy bin.  At these multipoles, no significant angular power is seen in the model in any energy bin.  The excess power measured in the data at these multipoles suggests a contribution from a point source population not present in the model.  At multipoles $\ell \lesssim 100$, angular power above the noise is clearly seen in both the data and model up to energies of 10 GeV, and is likely due to contamination from the Galactic diffuse emission.

Above 10 GeV, no significant power above the photon noise level is measured in the data or the model at any multipole.  However, we emphasize that due to decreasing photon statistics, the amplitude of anisotropies detectable by this analysis decreases with increasing energy, hence the measurements at higher energies currently do not exclude the presence of anisotropies at those energies at the level detected at 1 - 10 GeV\@.

Prior theoretical work on anisotropies in the IGRB has generated predictions for the fluctuation angular power spectra of many potential contributors to the measured IGRB emission.  Although the predictions for a given source class can vary by orders of magnitude depending on the assumptions made, some representative values of $C_{\ell}$ at $\ell = 100$ include $\sim 10^{-4}$ sr for blazars~\cite{Ando:2006cr}, $\sim 10^{-7}$ sr for star-forming galaxies \cite{Ando:2009nk}, and between $\sim 10^{-4}$ sr and $\sim 0.1$ sr for dark matter annihilation in Galactic subhalos \cite{Ando:2009fp}.  In the energy range of 1 - 5 GeV, for multipoles $\ell \gtrsim 100$, the value $C_{\ell} \simeq 10^{-5}$ sr measured in this analysis falls generally in the range predicted for some astrophysical source classes and some dark matter scenarios for the angular power spectrum of the emission from a single source class.  This result suggests that angular power spectrum measurements using \emph{Fermi}-LAT data may provide a sensitive probe of anisotropy signatures from known and proposed gamma-ray source classes.

\acknowledgments
The \textit{Fermi} LAT Collaboration acknowledges generous ongoing support
from a number of agencies and institutes that have supported both the
development and the operation of the LAT as well as scientific data analysis.
These include the National Aeronautics and Space Administration and the
Department of Energy in the United States, the Commissariat \`a l'Energie Atomique
and the Centre National de la Recherche Scientifique / Institut National de Physique
Nucl\'eaire et de Physique des Particules in France, the Agenzia Spaziale Italiana
and the Istituto Nazionale di Fisica Nucleare in Italy, the Ministry of Education,
Culture, Sports, Science and Technology (MEXT), High Energy Accelerator Research
Organization (KEK) and Japan Aerospace Exploration Agency (JAXA) in Japan, and
the K.~A.~Wallenberg Foundation, the Swedish Research Council and the
Swedish National Space Board in Sweden.  Additional support for science analysis during the operations phase is gratefully acknowledged from the Istituto Nazionale di Astrofisica in Italy and the Centre National d'\'Etudes Spatiales in France.

\end{document}